\documentclass[%
 aps,      
 pre,      
 amsmath,amssymb,
 reprint,%
 onecolumn,
 groupedaddress,
]{revtex4-1}

\usepackage{graphicx}  
\usepackage{dcolumn}   
\usepackage{bm}        
\usepackage{verbatim}
\usepackage{amsmath}

\begin{document}

\title[Self-Limiting Excitation of MEMS Devices with Surface Electrodes]{Self-Limiting Excitation of MEMS Devices with Surface Electrodes}

\author{Pavel E. Kornilovitch}
\affiliation{Hewlett-Packard Company, Printing and Personal Systems, Corvallis, Oregon 97330 USA} 
\email{pavel.kornilovich@hp.com}

\author{Robert G. Walmsley}
\affiliation{Hewlett-Packard Laboratories, Palo Alto, California 94304 USA} 
\email{bob.walmsley@hp.com}

\date{\today}  

\begin{abstract}

An excitation method for MEMS devices with planar electrodes is described. The stationary part of the device (the stator) consists of three electrode arrays arranged in the $ABCABC$ order. $A$, $B$, and $C$ carry time-independent potentials and together form a spatially-periodic electrostatic profile. The moving part of the device (the translator) has two electrode arrays $ababab$, with $a$ and $b$ carrying time-dependent out-of-phase voltages. When the frequency of the time-dependent voltage is close to the natural frequency of the spring-mass system, the translator is driven into resonance. By adjusting the spatial phase of the stationary profile, the driving force on the translator can be maximized for any equilibrium position. Physical misalignment between the stator and translator resulting from imperfect fabrication can be corrected electrically. A dynamical equation describing translator motion is derived and analyzed for resonant and parametric driving. In both cases, the driving force depends on the translator displacement in a periodic fashion. Such nonlinearity of the driving force results in self-stabilization of forced oscillations. This property has implications for the stability of vibratory gyroscopes.         

\end{abstract}


\maketitle

\section{\label{sec:zero}
Introduction
}

In vibratory MEMS gyroscopes proof masses are driven into resonance to maximize the amplitude of a Coriolis force~\cite{Shkel2009}. Amplitude or frequency variations of this motion directly translate into variations of the Coriolis signal. Gyroscopes with low angle random walk require stable primary oscillations insensitive to temperature, pressure and other environmental changes. One way to excite the proof mass is by using comb drives~\cite{Bernstein1993,Tanaka1995,Oh1997,Funk1999,Geen2002,Xie2003,Sharma2007}. In this method, the driving force is independent of the proof mass position and the oscillation amplitude is determined by the balance between external, elastic and dissipation forces. Changes in the quality factor $Q$ due to environmental variations directly affect the amplitude, which in turn causes the scale factor to drift. In practical devices, these effects are typically mitigated by an amplitude control loop~\cite{Shkel2009}.        

The intrinsic $Q$-sensitivity of the resonator can be reduced in nonlinear drives if the driving force amplitude decreases with displacement. (Note that in common gap-closing drives the force increases with displacement.) An example of such a system would be the magnetic pendulum in crossed steady and oscillating magnetic fields~\cite{Moon1987} or the gravitational pendulum driven by a horizontal harmonic force~\cite{Jeong1999}. Since only the tangential component of the external force affects the motion, the driving force decreases away from the equilibrium position. As the resonator is excited, energy transfer slows down and approaches zero when a stationary amplitude is reached. As long as dissipation is not dominant, the final stationary amplitude will be determined by the restoring and external forces only, thus eliminating dependence on $Q$.         

In this paper, we describe another nonlinear driving method, referred to as three-phase driving, which has a similar property of self-stabilization. Unlike pendulum, the proof mass undergoes linear motion and is excited by planar electrodes deposited on two flat surfaces separated by a micron-scale gap. The electrode arrangement is similar to that of three-phase accelerometer~\cite{Homeijer2010}, but in this case electrostatic interaction between the stationary and moving electrodes are used to excite mechanical motion. The electrode geometry is defined in section~\ref{sec:one}, capacitance matrix analyzed in section~\ref{sec:three}, and electrostatic force derived in section~\ref{sec:four}. The nonlinear equations of motion are derived and solved in section~\ref{sec:six} for resonant excitation and in section~\ref{sec:seven} for parametric excitation.        

Another important feature of three-phase driving is misalignment tolerance. By adjusting the voltages of stationary electrodes one can always tune the maximum of the driving force to coincide with the mechanical equilibrium of the proof mass. Any misalignment between the moving and stationary parts caused by imperfect fabrication can be compensated electrically. This property is derived in section~\ref{sec:five}.

\section{\label{sec:one}
Electrode geometry 
}

\begin{figure}[t]
\begin{center}
\includegraphics[width=0.48\textwidth]{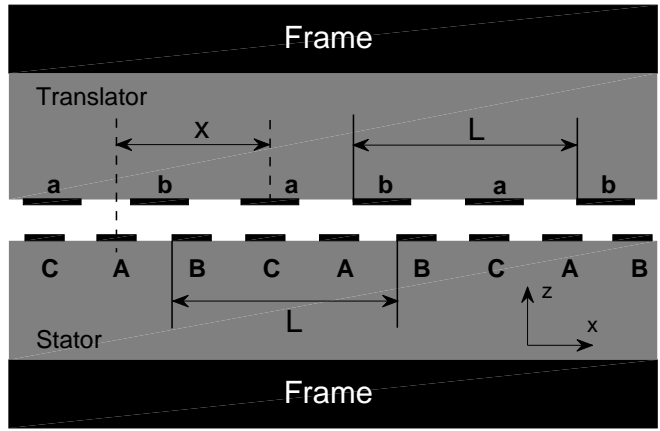}
\end{center}
\caption{The basic electrode geometry. $A$, $B$, and $C$ are stationary. The translator electrodes $a$ and $b$ can move along the $x$ direction. $x$ is referenced from the position in which the center of $a$ aligns with the center of $A$.  The electrodes are supported by dielectric substrates with a static dielectric constant of 12.1 (single-crystal silicon). The middle gap has a dielectric constant of 1.0. Notice the periodic boundary conditions along the $x$-axis.  Typical dimensions are (all in microns): array pitch $L = 24.0$, bottom dielectric thickness = 10.0; bottom electrodes width = 4.0; bottom electrodes height = 0.2; top dielectric thickness = 10.0; top electrode width = 6.0; top electrode height = 0.2; the stator-translator gap = 1.6.}
\label{fig:one}
\end{figure}

Both stationary and moving electrodes are periodic arrays of parallel strip lines extended along the $y$-axis. It is sufficient to consider only the two-dimensional $(xz)$ cross section of the system, which is shown in figure~\ref{fig:one}. The stationary electrodes are split into three different groups $A$, $B$, and $C$, and arranged in a periodic sequence $ABCABCABC\! \ldots$  Physically, all the electrodes are identical, with equal widths and equal gaps in between. The center-to-center distance between two nearest electrodes of the same group is $L$, which is the spatial period (pitch) of the array. The smallest center-to-center $AB$ distance (equal to the smallest $BC$ and $CA$ distances) is $L/3$. All electrodes that belong to the same group are electrically connected and carry the same time independent potential. The potentials are set according to the three-phase driving rule:
\begin{eqnarray}
\phi_{A} & = & V \cos{\left( \theta                  \right)} \: ,  
\label{eq:one} \\
\phi_{B} & = & V \cos{\left( \theta + \frac{2\pi}{3} \right)} \: ,  
\label{eq:two} \\
\phi_{C} & = & V \cos{\left( \theta + \frac{4\pi}{3} \right)}  
           =   V \cos{\left( \theta - \frac{2\pi}{3} \right)} \: .  
\label{eq:three} 
\end{eqnarray}
The amplitude $V$ and phase $\theta$ are two adjustable parameters. In practical devices, $V$ and $\theta$ are set by external electronics. Collectively the $A$, $B$, and $C$ electrodes will be referred to as the stator.

The moving electrodes are split into two groups, to be denoted $a$ and $b$, and arranged in the sequence $abababab\ldots$  Again, the electrodes are physically identical and equidistant. Their width and gap are in general different from those of $A$, $B$, and $C$, but the total spatial period is equal to $L$. The center-to-center distance of the nearest $ab$ pair is $L/2$. Collectively, the $ab$ electrode array will be referred to as the translator. In three-phase driving, $a$ and $b$ carry out-of-phase time-dependent voltages  
\begin{eqnarray}
\phi_{a}(t) & = & U \cos{\left( \omega t       \right)} \equiv U(t) \: ,  
\label{eq:four} \\
\phi_{b}(t) & = & U \cos{\left( \omega t + \pi \right)} = - U(t) \: .  
\label{eq:five}  
\end{eqnarray}
The amplitude $U$ and the angular frequency $\omega$ are set by external electronics. The overall phase of $U(t)$ is unimportant and will be set to zero. 

The rest of the device is represented by the mechanical support, which is electrically insulating, and by the frame electrode $f$. The latter is always electrically grounded, $\phi_f = 0$. The frame electrode is assumed to be uniform along the $x$ axis. 

To describe translator motion, a coordinate reference is needed. In this paper, $x = 0$ corresponds to the center of an $a$ electrode coinciding with the center of an $A$ electrode, as indicated in figure~\ref{fig:one}. Assuming linear elastic force and linear dissipation, and neglecting static friction and stochastic noise, the equation of motion for a translator mass $M$ reads 
\begin{equation}
M \ddot{x} + b \dot{x} + \kappa x = - \frac{\partial W(x)}{\partial x} \: . 
\label{eq:seven}
\end{equation}
Here $W(x)$ is the quasi-electrostatic energy of the system, whose $x$-derivative is the external force acting on the translator. An expression for $W(x)$ under the three-phase rules is derived in the next two sections.

\section{\label{sec:three}
Capacitance matrix 
}

The device under study is a quasi-stationary system of conductors with defined electrostatic potentials. The full electrostatic energy is given by
\begin{equation}
W(x) = \frac{1}{2} \sum^{N}_{i\, =1} q_i(x) \phi_i = 
\frac{1}{2} \sum^{N}_{i,j \, = 1} C_{i\! j}(x) \, \phi_i \phi_{\! j} \: , 
\label{eq:nine}
\end{equation}
where $N = 6$ is the total number of conductors. ($N = A, B, C, a, b, f$.) The symmetrical capacitance matrix $C_{i\! j}$ is defined as
\begin{equation}
q_i = \sum^{N}_{i=1} C_{i\! j} \, \phi_j \: . 
\label{eq:eight}
\end{equation}
In three-phase driving, the capacitance coefficients are functions of the translator position $x$. As a result, the electrode charges are also position-dependent, $q_i = q_i(x)$. 
 
In general, the energy (\ref{eq:nine}) comprises 21 different contributions and depends on $N(N+1)/2 = 21$ capacitance functions. Since the frame electrode is assumed to be always grounded, $\phi_f = 0$, the number of terms is reduced to 15. In addition, translation, reflection, and permutation  symmetries within the $A$, $B$, $C$, and $a$, $b$ groups reduce that number to just 8 irreducible functions. The symmetry properties are summarized in table~\ref{tab:one}.

\begin{table}
\renewcommand{\tabcolsep}{0.2cm}
\renewcommand{\arraystretch}{1.5}
\begin{center}
\begin{tabular}{|c|c|c|c|}
\hline\hline
 Function                  &  Spatial period  & Parity $(x \rightarrow -x)$ &  Generating function   \\ \hline
\hline 
 $C_{aa}(x)$               &      $L/3$       &  even                       &  $\equiv G_1(x)$       \\ \hline
 $C_{bb}(x)$               &      $L/3$       &  even                       &  $G_1(x+L/6)$          \\ \hline
\hline 
 $C_{ab}(x)$, $C_{ba}(x)$  &      $L/6$       &  even                       &  $\equiv G_2(x)$       \\ \hline
\hline
 $C_{af}(x)$, $C_{fa}(x)$  &      $L/3$       &  even                       &  $\equiv G_3(x)$       \\ \hline
 $C_{bf}(x)$, $C_{fb}(x)$  &      $L/3$       &  even                       &  $G_3(x+L/6)$          \\ \hline
\hline
 $C_{AA}(x)$               &      $L/2$       &  even                       &  $\equiv G_4(x)$       \\ \hline
 $C_{BB}(x)$               &      $L/2$       & $C_{BB}(x) = C_{CC}(-x)$    &  $G_4(x-L/3)$          \\ \hline
 $C_{CC}(x)$               &      $L/2$       & $C_{CC}(x) = C_{BB}(-x)$    &  $G_4(x+L/3)$          \\ \hline
\hline
 $C_{AB}(x)$, $C_{BA}(x)$  &      $L/2$       & $C_{AB}(x) = C_{AC}(-x)$    &  $G_5(x-L/6)$          \\ \hline
 $C_{AC}(x)$, $C_{CA}(x)$  &      $L/2$       & $C_{AC}(x) = C_{AB}(-x)$    &  $G_5(x+L/6)$          \\ \hline
 $C_{BC}(x)$, $C_{CB}(x)$  &      $L/2$       &  even                       &  $\equiv G_5(x)$       \\ \hline
\hline
 $C_{Af}(x)$, $C_{fA}(x)$  &      $L/2$       &  even                       &  $\equiv G_6(x)$       \\ \hline
 $C_{Bf}(x)$, $C_{fB}(x)$  &      $L/2$       & $C_{Bf}(x) = C_{Cf}(-x)$    &  $G_6(x+L/6)$          \\ \hline
 $C_{Cf}(x)$, $C_{fC}(x)$  &      $L/2$       & $C_{Cf}(x) = C_{Bf}(-x)$    &  $G_6(x-L/6)$          \\ \hline
\hline
 $C_{aA}(x)$, $C_{Aa}(x)$  &      $L$         &  even                       &  $\equiv G_7(x)$       \\ \hline
 $C_{aB}(x)$, $C_{Ba}(x)$  &      $L$         & $C_{aB}(x) = C_{aC}(-x)$    &  $G_7(x-L/3)$          \\ \hline
 $C_{aC}(x)$, $C_{Ca}(x)$  &      $L$         & $C_{aC}(x) = C_{aB}(-x)$    &  $G_7(x+L/3)$          \\ \hline
 $C_{bA}(x)$, $C_{Ab}(x)$  &      $L$         &  even                       &  $G_7(x+L/2)$          \\ \hline
 $C_{bB}(x)$, $C_{Bb}(x)$  &      $L$         & $C_{bB}(x) = C_{bC}(-x)$    &  $G_7(x+L/6)$          \\ \hline
 $C_{bC}(x)$, $C_{Cb}(x)$  &      $L$         & $C_{bC}(x) = C_{bB}(-x)$    &  $G_7(x-L/6)$          \\ \hline
\hline 
 $C_{f\! f}(x)$            &      $L/6$       &  even                       &  $\equiv G_8(x)$       \\ \hline
\hline 
\end{tabular}
\end{center}
\caption{
Symmetry properties of the capacitance functions. ``Even'' means $C_{i \! j}(-x) = C_{i \! j}(x)$. ``$\equiv$'' means this relationship is chosen as the definition of the generating function in question. For example, $G_7(x)$ is defined as $C_{aA}(x)$. 
} 
\label{tab:one}
\end{table}

\begin{figure}[t]
\begin{center}
\includegraphics[width=0.98\textwidth]{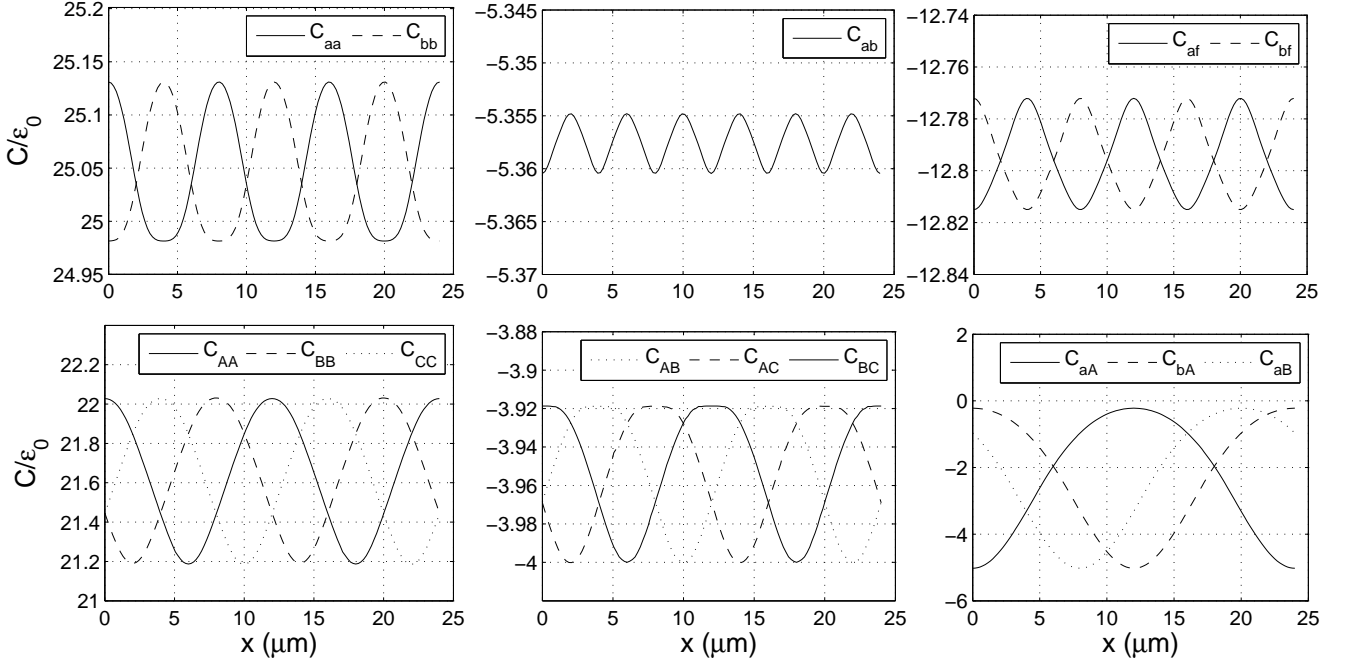}
\end{center}
\caption{The capacitance functions $C_{i\! j}(x)$ for the array pitch $L = 24$ $\mu$m and stator-translator gap of 1.6 $\mu$m. Other parameters are the same as in Fig.~\ref{fig:one}. The capacitances are per unit cell of the array, per unit length in the $y$ direction (1 m), and given in units of $\varepsilon_0 = 8.854 \cdot 10^{-12}$ F/m. Note different scales in different panels. The translation and permutation symmetries should be compared with table~\ref{tab:one}.}
\label{fig:onetwo}
\end{figure}

The capacitance coefficients $C_{i \! j}$ can be calculated numerically using finite-element software such as COMSOL Multiphysics. Following the definition (\ref{eq:eight}), the potential of one conductor is set to 1 V and the others are grounded. Then a two-dimensional Laplace equation is solved and the charge induced on all the conductors is calculated. Notice that the solution domain can be reduced to one irreducible unit cell $0 \leq x \leq L$ by imposing periodic boundary condition $\varphi(L,z) = \varphi(0,z)$ for all $z$. The procedure is repeated for multiple translator shifts $x$ to construct the entire capacitance functions. Several exemplary functions are shown in figure~\ref{fig:onetwo}. The electrode array pitch is $L = 24$ $\mu$m and the stator-translator gap is 1.6 $\mu$m. Other parameters are specified in the caption of figure~\ref{fig:one}. It can be observed that the functions shown in figure~\ref{fig:onetwo} satisfy the translation and permutation symmetries listed in table~\ref{tab:one}.

All eight generating functions are periodic and even. As such, they can be expanded in cosine Fourier series 
\begin{equation}
G_{i}(x) = \sum^{\infty \;\; \prime}_{n = 0, 1, \ldots} G^{(i)}_{n} \cos{ k_n \, x } \: , 
\label{eq:ten}
\end{equation}
where $k_n \equiv 2 \pi n/L$, and the prime at the sum sign means the $n=0$ term is taken with weight 1/2. All Fourier coefficients $G^{(i)}_n$ are real numbers and have the same dimensionality as the original $G_{i}(x)$. If the function itself is known then $G^{(i)}_n$ can be found by numerical integration. The Fourier coefficients for the parameters of figure~\ref{fig:one} and functions of figure~\ref{fig:onetwo} are summarized in table~\ref{tab:two}. The coefficients decay rapidly with $n$.

\begin{table}
\renewcommand{\tabcolsep}{0.2cm}
\renewcommand{\arraystretch}{1.5}
\begin{center}
\begin{tabular}{|c||c|c|c|c|}
\hline\hline
 Function   & $G^{(1)}_0$ & $G^{(1)}_3$ & $G^{(1)}_6$    & $G^{(1)}_9$     \\ \hline
 $G_1$      &  50.09      &  0.0777     &  0.0106         & -0.0029        \\ \hline
\hline
 Function   & $G^{(2)}_0$ & $G^{(2)}_6$ & $G^{(2)}_{12}$ & $G^{(i)}_{18}$  \\ \hline
 $G_2$      & -10.72      & -0.0027     &  0             &  0              \\ \hline
\hline
 Function   & $G^{(3)}_0$ & $G^{(3)}_3$ & $G^{(3)}_{6}$  & $G^{(3)}_{9}$   \\ \hline 
 $G_3$      & -25.59      & -0.0204     &  0.0013        &  0              \\ \hline
\hline
 Function   & $G^{(i)}_0$ & $G^{(i)}_2$ & $G^{(i)}_4$    & $G^{(i)}_6$     \\ \hline
 $G_4$      &  43.27      &  0.4163     & -0.0222        &  0.0053         \\ \hline
 $G_5$      &  -7.90      &  0.0411     & -0.0069        & -0.0003         \\ \hline
 $G_6$      & -18.27      &  0.2789     &  0.0057        &  0.0010         \\ \hline
\hline
 Function   & $G^{(7)}_0$ & $G^{(7)}_1$ & $G^{(7)}_2$    & $G^{(7)}_3$     \\ \hline
 $G_7$      & -4.5957     & -2.3828     & -0.3271        & -0.0189         \\ \hline
\hline
 Function   & $G^{(8)}_0$ & $G^{(8)}_6$ & $G^{(8)}_{12}$ & $G^{(8)}_{18}$  \\ \hline 
 $G_8$      & 105.99      & -0.0055     &  0             &  0              \\ \hline
\hline  
\end{tabular}
\end{center}
\caption{
First four non-vanishing Fourier coefficients for the eight generating functions. The values are given in units of $\varepsilon_0$. The geometry parameters are given in the caption of figure~\ref{fig:one}. The zero-order coefficients of the frame functions $G^{(3)}_0$, $G^{(6)}_0$ and $G^{(8)}_0$ are unphysically large because the dielectric thicknesses are unrealistically small (10 $\mu$m). These coefficients do not affect the electrostatic force on the translator.  
} 
\label{tab:two}
\end{table}

The dynamics of the translator is for the most part determined by the generating function $G_7(x)$, see sec\-tion~\ref{sec:four}. This function is smooth and its Fourier expansion converges very rapidly with the number of harmonics included. Already the first three coefficients reconstruct the function with a better than 1\% accuracy. Five coefficients are sufficient to accurately represent $G_7(x)$ at all $x$.

\section{\label{sec:four}
Electrostatic energy and force
}

Referring to equation~(\ref{eq:nine}), the total number of energy terms is 21. However, since $\phi_f = 0$, only 15 terms remain. Next, one makes use of the three-phase rules (\ref{eq:four}) and (\ref{eq:five}) implying that the $a$ and $b$ potentials are always out-of-phase. Expanding the double sum, one obtains
\begin{eqnarray}
W & = & \frac{1}{2} \, U^2(t) \left( C_{aa} -2 C_{ab} + C_{bb} \right) 
\nonumber \\
  &  & + \frac{1}{2} \left( C_{AA} \phi^2_{A} + C_{BB} \phi^2_{B} + C_{CC} \phi^2_{C} \right) 
    +  \left( C_{AB} \phi_{A} \phi_{B} + C_{AC} \phi_{A} \phi_{C} + C_{BC} \phi_{B} \phi_{C} \right)
\nonumber \\
  &  & + U(t) \left[ \phi_{A} ( C_{aA} - C_{bA} ) + \phi_{B} ( C_{aB} - C_{bB} )   
                   + \phi_{C} ( C_{aC} - C_{bC} ) \right]  \: .  
\label{eq:twentyfive} 
\end{eqnarray}
The first line here involves only the translator electrodes and therefore can be interpreted as a translator self-energy. The second line involves only the stationary electrodes and can be interpreted as a stator self-energy. The last line is a stator-translator interaction energy. Note that the translator and stator self-energies are in general functions of $x$ and as such may contribute to the dynamics.    

The capacitance functions are now expressed via five generating functions according to the rules of table~\ref{tab:one}, and the stator potentials are substituted from equations~(\ref{eq:one})-(\ref{eq:three}). The result is
\begin{eqnarray}
W(x) & = & W_{\rm tr}(x) + W_{\rm st \: diag}(x) + W_{\rm st \: off-diag}(x) + W_{\rm int}(x)  
\nonumber \\
  & = & \frac{1}{2} \, U^2(t) \left[ G_{1}(x) + G_1(x+L/6) - 2 G_{2}(x) \right] 
\nonumber \\
  & + & \frac{1}{2} \, V^2 \left[ G_{4}(x)     \cos^2{\left( \theta          \right) } + 
                                  G_{4}(x-L/3) \cos^2{\left( \theta + 2\pi/3 \right) } \right. 
\nonumber \\
  &   & \makebox[3.0cm]{} \left. + \: G_{4}(x+L/3) \cos^2{\left( \theta - 2\pi/3 \right) }   \right]
\nonumber \\
  & + & V^2 \left[ G_5(x)     \cos{\left( \theta + 2\pi/3 \right) } 
                              \cos{\left( \theta - 2\pi/3 \right) }  \right. 
\nonumber \\
  &   &  \makebox[3.0cm]{}  +  G_5(x-L/6) \cos{\left( \theta \right) } \cos{\left( \theta + 2\pi/3 \right) } 
\nonumber \\                  
  &   &  \makebox[3.0cm]{}  \left. + \: G_5(x+L/6) \cos{\left( \theta \right) } \cos{\left( \theta - 2\pi/3 \right) }
            \right]
\nonumber \\
  & + & V U(t) \left\{ \cos{\left( \theta          \right) } \left[ G_7(x)     - G_7(x+L/2) \right] \right.
\nonumber \\  
  &   &  \makebox[3.0cm]{}   + \cos{\left( \theta + 2\pi/3 \right) } \left[ G_7(x-L/3) - G_7(x+L/6) \right]    
\nonumber \\  
  &   &  \makebox[3.0cm]{} \left. + \cos{\left( \theta - 2\pi/3 \right) } \left[ G_7(x+L/3) - G_7(x-L/6) 
                                                                                                \right] \right\}  .  
\label{eq:twentysix} 
\end{eqnarray}
Next, Fourier expansions should be substituted in place of the capacitance functions and the resulting expressions simplified. The four contributions will now be considered separately. 
\newline
{\em Translator self-energy.} When summing the Fourier series for $G_1(x)$, odd terms cancel while even terms double. As a result, the translator self-energy assumes the form
\begin{equation}
W_{\rm tr}(x) = U^2(t)  
\sum^{\infty \;\; \prime}_{n=0} \left[ G^{(1)}_{6n} - G^{(2)}_{6n} \right] \cos{ k_{6n} x }  \: .
\label{eq:twentyseven}
\end{equation}
\newline
{\em Diagonal stator self-energy.} The diagonal stator self-energy involves generating function $G_4(x)$. Upon substitution of the corresponding Fourier expansion, it is convenient to consider three sets of harmonics separately: (i) $0, 6, 12, \ldots = 6n$; (ii) $2, 8, 14, \ldots = 2 + 6n$, and (iii) $4, 10, 16, \ldots = 4 + 6n$. In all three cases, a combination of trigonometric functions is simplified to a concise expression. The final result is 
\begin{eqnarray}
W_{\rm st \: diag}(x) & = & \frac{3}{8} \, V^2 \left\{ 
2 \sum^{\infty \;\; \prime}_{n=0} G^{(4)}_{6n} \cos{ k_{6n} x } \right.
\nonumber \\
&  & + \left. \sum^{\infty}_{n=0} \left[ G^{(4)}_{2+6n} \cos{ ( k_{2+6n} x + 2 \theta ) } + 
                                         G^{(4)}_{4+6n} \cos{ ( k_{4+6n} x - 2 \theta ) }   \right] 
\right\}   . 
\label{eq:twentyeight}
\end{eqnarray}
\newline
{\em Off-diagonal stator self-energy.} Again, three sets of $G_5(x)$ harmonics are treated separately. After some algebra, the off-diagonal stator self-energy is
\begin{eqnarray}
W_{\rm st \: off-diag}(x) & = & \frac{3}{4} \, V^2 \left\{ 
- \sum^{\infty \;\; \prime}_{n=0} G^{(5)}_{6n} \cos{ k_{6n} x } \right. 
\nonumber \\
& & + \left. \sum^{\infty}_{n=0} \left[ G^{(5)}_{2+6n} \cos{ ( k_{2+6n} x + 2 \theta ) } + 
                                        G^{(5)}_{4+6n} \cos{ ( k_{4+6n} x - 2 \theta ) }    \right] 
\right\}   .
\label{eq:twentynine}
\end{eqnarray}
The structure of equation~(\ref{eq:twentynine}) is similar to that of the diagonal stator energy, equation~(\ref{eq:twentyeight}).
\newline
{\em Stator-translator interaction energy.} In the last term of equation~(\ref{eq:twentysix}) harmonics with even $n$ cancel while those with odd $n$ double. The interaction energy becomes
\begin{eqnarray}
W_{\rm int}(x) & = & 2 V U(t) \sum^{\infty}_{n = 1, 3, \ldots} G^{(7)}_n
\left[ \cos{k_n x}            \cos{\left( \theta \right) }           \right.
\nonumber \\
  &   &  \makebox[3.0cm]{} + \cos{(k_n x - 2\pi n/3)} \cos{\left( \theta + 2\pi/3 \right) } 
\nonumber \\  
  &   &  \makebox[3.0cm]{} + \left. \cos{(k_n x + 2\pi n/3)} \cos{\left( \theta - 2\pi/3 \right) } \right] \: .  
\label{eq:thirty} 
\end{eqnarray}
Transformations are convenient to perform separately for the following groups of harmonics: $1, 7, 13, \ldots = 1 + 6n$, $3, 9, 15, \ldots = 3 + 6n$, and $5, 11, 17, \ldots = 5 + 6n$. The final result is 
\begin{equation}
W_{\rm int}(x) = 3 V U(t) \sum^{\infty}_{n = 0} \left\{ 
G^{(7)}_{1+6n} \cos{ (k_{1+6n} x + \theta) } +
G^{(7)}_{5+6n} \cos{ (k_{5+6n} x - \theta) }    \right\} .  
\label{eq:thirtyone} 
\end{equation}
Note that harmonics with $n = 3, 9, 15, \ldots$ have vanished. This is a consequence of three-phase rules (\ref{eq:one})-(\ref{eq:three}).    

Total electrostatic energy is given by the sum of equations (\ref{eq:twentyseven}), (\ref{eq:twentyeight}), (\ref{eq:twentynine}) and (\ref{eq:thirtyone}). The electrostatic force can be found as a negative derivative of the energy with respect to translator position $x$. Performing differentiation one finds (recall that $k_n \equiv 2\pi n/L$):
\begin{eqnarray}
F_{\rm el-st}(x) & = & U^2(t) \sum^{\infty}_{n=1} 
k_{6n} \left[ G^{(1)}_{6n} - G^{(2)}_{6n} \right] \sin{ k_{6n} x } 
+ \frac{3}{4} \, V^2 \sum^{\infty}_{n=1} k_{6n} \left[ G^{(4)}_{6n} - G^{(5)}_{6n} \right] \sin{ k_{6n} x }
\nonumber \\
 &   & + \frac{3}{8} \, V^2 \sum^{\infty}_{n=0} \left\{ 
 k_{2+6n} \left[ G^{(4)}_{2+6n} + 2 G^{(5)}_{2+6n} \right] \sin{ ( k_{2+6n} x + 2 \theta ) } \right.
\nonumber \\ 
 &   & \makebox[2.0cm]{} \left. + \: k_{4+6n} \left[ G^{(4)}_{4+6n} + 2 G^{(5)}_{4+6n} \right] 
 \sin{ ( k_{4+6n} x - 2 \theta ) }  \right\} 
\nonumber \\
 &   & + 3 V U(t) \sum^{\infty}_{n = 0} \left\{ 
k_{1+6n} G^{(7)}_{1+6n} \sin{ (k_{1+6n} x + \theta) }  \right. 
\nonumber \\
 &   & \makebox[2.0cm]{} \left. + \: k_{5+6n} G^{(7)}_{5+6n} \sin{ (k_{5+6n} x - \theta) }    \right\}  \: .
\label{eq:thirtytwo}
\end{eqnarray}
Formula (\ref{eq:thirtytwo}) is general. Based on the results of finite-element modelling summarized in table~\ref{tab:two}, the general expression can be reduced to a handful of the most relevant terms. The following observations are made. (i) The translator self-force begins with coefficients $C^{(1)}_6$ and $C^{(2)}_6$. According to table~\ref{tab:two}, both coefficients are of the order 0.01 or less, and hence can be neglected. The translator self-force does not contribute to its own dynamics. (ii) The stator self-force contains even harmonics of functions $G_4(x)$ and $G_5(x)$. Only the $n=2$ coefficients are larger than 0.04; the rest can be omitted. The stator self-force does contribute to the translator dynamics. (iii) The interaction force starts with coefficients $G^{(7)}_1$ and $G^{(7)}_5$. The latter is of the order 0.0003, and should be omitted. Thus although the main driving function $G_7(x)$ is highly anharmonic, its effect is represented by the sole coefficient $C^{(7)}_1$ with a very high degree of accuracy. Putting all of this together, one arrives at a truncated electrostatic force in the following form:
\begin{eqnarray}
\tilde{F}_{\rm el-st}(x) & = & \frac{3}{8} \, V^2 
\left\{ k_{2}  \left[ G^{(4)}_{2} + 2 G^{(5)}_{2} \right] \sin{ ( k_{2} x + 2 \theta ) } \right\} 
\nonumber \\
& & \makebox[2.0cm]{} + 3 V U \cos{ \omega t } \left\{ k_{1} G^{(7)}_{1} \sin{ (k_{1} x + \theta) } \right\}  .
\label{eq:thirtythree}
\end{eqnarray}
This expression can be used for all practical purposes. Including other small terms would have exceeded the accuracy of the numerical solution and of the model itself. The first term in the force originates from the stator self-energy. It is present even if translator electrodes are grounded. This force has no explicit time dependence and can be thought of as a correction to the elastic force acting on the translator. This correction is proportional to the square of the three-phase stator amplitude $V^2$. The second term in equation~(\ref{eq:thirtythree}) originates from the interaction between the stator and the translator. It oscillates in time with a frequency $\omega$ and is the main driving force of the device. This force is nonzero only if both the stator and translator are at potentials prescribed by the three-phase rules.

\section{\label{sec:five}
Selection of $\theta$
}

Choosing optimal $\theta$ is now discussed. In the presence of stator-translator misalignment, the translator is permanently shifted by a distance $x_0$ from the designed position. The elastic energy of a linear spring is 
\begin{equation}
W_{\rm elastic}(x) = \frac{\kappa}{2} \left( x - x_0 \right)^2 = \frac{\kappa \, \bar{x}^2}{2} \: ,  
\label{eq:thirtyfive} 
\end{equation}
where $\bar{x} = x - x_0$ is the new independent variable. Since time derivatives of $\bar{x}$ are the same as of $x$, equation of motion (\ref{eq:seven}) becomes
\begin{equation}
M \ddot{\bar{x}} + b \dot{\bar{x}} + \kappa \, \bar{x} = 
\tilde{F}_{\rm el-st}( \bar{x} + x_0 ) \: . 
\label{eq:thirtyeight}
\end{equation}
Consider now the main driving force, which is the second term in equation~(\ref{eq:thirtythree}). Its $x$-dependence is defined by the factor  
\begin{equation}
\sin{ (k_{1} x + \theta) } = \sin{ ( k_{1} \bar{x} + k_{1} x_0 + \theta ) } \: . 
\label{eq:thirtynine}
\end{equation}
In forced oscillations, it is desirable that the external force is maximal at the equilibrium position $\bar{x} = 0$ where the velocity is maximal. This condition maximizes the mechanical power absorbed by the translator. Expression~(\ref{eq:thirtynine}) is maximized by choosing $\theta$ such that 
\begin{equation}
k_{1} x_0 + \theta = \frac{\pi}{2} \: . 
\label{eq:forty}
\end{equation}
Then the driving force 
\begin{equation}
3 k_1 V U G^{(7)}_{1} \cos{ ( k_{1} \bar{x} ) } \cos{ \omega t }  
\label{eq:fortyone}
\end{equation}
is maximal at $\bar{x} = 0$, as desired. The relation (\ref{eq:forty}) expresses an important property of three-phase driving. By adjusting three phase $\theta$ electronically one can always cancel the effects of physical misalignment, correct for process variations, and maximize the driving force. 
 
The stator-stator induced force is defined by the following factor [the first term in equation~(\ref{eq:thirtythree}); note that $k_2 = 2 k_1$]:
\begin{equation}
\sin{ ( k_{2} x + 2 \theta ) }  = \sin{ [ k_{2} \bar{x} + 2 ( k_1 x_0 + \theta ) ] } = 
\sin{ ( k_{2} \bar{x} + \pi ) } = - \sin{ ( k_{2} \bar{x} ) } \: .  
\label{eq:fortytwo}
\end{equation}
Collecting all the terms one obtains the final equation of motion
\begin{equation}
 M \ddot{\bar{x}} + b \dot{\bar{x}} + \kappa \, \bar{x}  
+ \frac{3}{8} \, k_2 V^2 \! \left\{ G^{(4)}_{2} \!\! + 2 G^{(5)}_{2} \! \right\} \sin{ ( k_{2} \bar{x} ) } =
3 \, k_1 V U G^{(7)}_{1} \cos{ ( k_{1} \bar{x} ) } \cos{ \omega t } \: ,
\label{eq:fortythree}
\end{equation}
which is a central result of the paper. The last term on the left does not depend explicitly on time, and is interpreted as an electrostatic contribution to the translator restoring force. Note that it has the same symmetry as the elastic force: it is odd in $\bar{x}$. At small vibration amplitudes the electrostatic force simply renormalizes spring stiffness $\kappa$. At large amplitudes, it brings about new effects. Both the electrostatic force and the driving force are periodic functions of the translator displacement $\bar{x}$. Thus the equation of motion (\ref{eq:fortythree}) is highly nonlinear. 

To facilitate subsequent analysis, it is convenient to rewrite the equation of motion in scaled variables. (i) Introduce dimensionless displacement $\xi$ and time $\tau$:
\begin{equation}
\xi \equiv k_1 \bar{x} = \frac{2 \pi \bar{x}}{L}   \: , \hspace{0.5cm} 
\tau \equiv \omega_0 t = t \sqrt{\frac{\kappa}{M}} \: , 
\label{eq:fortyfive}
\end{equation}
where $\omega_0 = \sqrt{\kappa/M}$ is the translator's natural frequency at small amplitudes. Hereafter, the derivative with respect to $\tau$ will be denoted by a prime rather than a dot. (ii) Divide the equation of motion by $( M \omega^2_0 L )/(2\pi)$. (iii) Introduce new dimensionless parameters
\begin{equation}
\epsilon \equiv \frac{b}{2 M \omega_0} \: ,
\label{eq:fortyeight}
\end{equation}
\begin{equation}
\delta \equiv \frac{3 \, V^2}{4 \, M\omega^2_0} \left( \frac{2\pi}{L} \right)^2 
\left\{ G^{(4)}_{2} + 2 G^{(5)}_{2} \right\} , 
\label{eq:fifty}
\end{equation}
\begin{equation}
g \equiv \frac{3 V U G^{(7)}_{1}}{M\omega^2_0} \left( \frac{2\pi}{L} \right)^2 , 
\label{eq:fiftyone}
\end{equation}
\begin{equation}
\Omega \equiv \frac{\omega}{\omega_0} \: . 
\label{eq:fiftytwo}
\end{equation}
The equation of motion~(\ref{eq:fortythree}) becomes
\begin{equation}
\xi'' + 2 \epsilon \xi' + \xi   
+ \delta \sin{ ( 2\xi ) } = g \cos{ ( \xi ) } \cos{ \Omega \tau }  . 
\label{eq:fiftythree}
\end{equation}
The four parameters have the following meaning. (i) $\epsilon = 1/(2Q)$ determines the rate of energy dissipation in the system; it is an inverse of the quality factor $Q$. (ii) $\delta$ is the measure of electrostatic force; it is proportional to $V^2$; (iii) $g$ is the dimensionless amplitude of the main driving force; (iv) $\Omega$ is the dimensionless external frequency that characterizes frequency mismatch between the driving force and the natural frequency. Typical parameter values are given in table~\ref{tab:three}.

\begin{table}
\renewcommand{\tabcolsep}{0.1cm}
\renewcommand{\arraystretch}{1.5}
\begin{center}
\begin{tabular}{|c||c|c|c|c|}
\hline\hline
 Parameter  & $\epsilon$ & $\delta$ &   $g$    &      $\Omega$       \\ \hline
 Value      &  0.00076$^{\rm \, a}$  
                         &  0.0037$^{\rm \, b}$  
                                    &  0.036$^{\rm \, c}$  
                     & $\approx 1^{\rm \, d}$ or $\approx 2^{\rm \, e}$    \\ \hline
\hline
\end{tabular}
\end{center}
\caption{
Typical values of the dimensionless parameters (\ref{eq:fortyeight})-(\ref{eq:fiftytwo}) for the geometry of figure~\ref{fig:one}. The estimates assume the translator wafer thickness 125 $\mu$m, silicon mass density $2330$ kg/m$^3$, natural frequency $\omega_0 = 2\pi \cdot 6500$ 1/s, stator voltage amplitude $V = 8$ volts, and translator carrier amplitude $U = 12$ volts. $^{\rm a}$Estimated from ringdown data. $Q = 1/(2\epsilon) = 658$. $^{\rm b}$Estimated from equation~(\ref{eq:fifty}) and table~\ref{tab:two}. $^{\rm c}$Estimated from equation~(\ref{eq:fiftyone}) and table~\ref{tab:two}. $^{\rm d}$Resonant driving. $^{\rm e}$Parametric driving.  
} 
\label{tab:three}
\end{table}

\section{\label{sec:six}
Resonant driving
}

The most interesting feature of the equation of motion (\ref{eq:fiftythree}) is the nonlinear driving force $g \cos{(\xi)} \cos{\Omega t}$. Nonlinearity of such type have been studied in relation to cha\-o\-tic behaviours of driven magnetic pendulum~\cite{Moon1987}, horizontally driven gravitational pen\-du\-lum~\cite{Jeong1999}, elliptically driven gravitational pendulum~\cite{Fidlin2008,Horton2011,Pavlovskaia2012}, and charged particles in the field of plane waves~\cite{Zaslavskii1972,Escande1981,Bialek1985}. In those physical systems, the functional form of the restoring force is linked to the functional form of drive nonlinearity. For example, the restoring force of the gravitational pendulum is proportional to $\sin{\xi}$ and has the same origin as the $\cos{\xi}$ of the driving force. In the system under study, the electrostatic correction $\delta \sin{2\xi}$ is similarly related to the driving force. However, the other part of the restoring force, $\xi$, comes from elastic flexures, i.e., from a different part of the system. In fact, other terms can be added to the restoring force, for example the Duffing term $\xi^3$, without any need to modify the driving force. The independence of restoring and driving forces adds to the richness of this dynamical system.    

The primary purpose of the present work is to study the nonlinear dynamics at small $g$ when oscillations are stable and no chaotic motion is observed. In addition, the electrostatic force will be neglected in the following analysis, despite its definite presence in practical devices. Setting $\delta = 0$, the equation of motion (\ref{eq:fiftythree}) reduces to  
\begin{equation}
\xi'' + 2 \epsilon \xi' + \xi = g \cos{ ( \xi ) } \cos{ \Omega \tau } \: . 
\label{eq:fiftyfive}
\end{equation}
In this section, resonant driving with $\Omega \approx 1$ will be considered. The goal is to derive resonance curves by applying the method of slowly changing amplitudes. A solution to equation~(\ref{eq:fiftyfive}) is sought in the following form
\begin{equation}
\xi(\tau) = \sigma(\tau) \cos{ \left[ \Omega \tau + \psi(\tau) \right] } \: . 
\label{eq:fiftysix}
\end{equation}
In the stationary regime, the amplitude $\sigma$ and phase $\psi$ are time-independent. Substitution in equation~(\ref{eq:fiftyfive}) yields
\begin{eqnarray}
\left( 1 - \Omega^2 \right) \sigma \cos{ \left( \Omega \tau + \psi \right) }
 - 2 \epsilon \, \Omega \sigma \sin{ \left( \Omega \tau + \psi \right) } = 
 \nonumber \\
\makebox[1.0cm]{} = g \cos{ \left( \Omega \tau \right) }  
 \cos{ \left\{ \sigma \cos{ \left( \Omega \tau + \psi \right) } \right\} } \: . 
\label{eq:fiftyseven}
\end{eqnarray}
Shifting the time origin by $\psi/\Omega$, the equation is brought to the form
\begin{eqnarray}
\left( 1 - \Omega^2 \right) \sigma \cos{ \left( \Omega \tau \right) }
     - 2 \epsilon \, \Omega \sigma \sin{ \left( \Omega \tau \right) } =
 \nonumber \\
\makebox[1.0cm]{} = g \cos{ \left( \Omega \tau \right) } \cos{ \psi } 
 \cos{ \left\{ \sigma \cos{ \left( \Omega \tau \right) } \right\} } 
 + g \sin{ \left( \Omega \tau \right) } \sin{ \psi } 
 \cos{ \left\{ \sigma \cos{ \left( \Omega \tau \right) } \right\} }  \: . 
\label{eq:fiftyeight}
\end{eqnarray}
By multiplying with $\cos{ \left( \Omega \tau \right) }$ and $\sin{ \left( \Omega \tau \right) }$ and averaging over the oscillation period $2\pi/\Omega$, one obtains a pair of equations
\begin{eqnarray}
\left( 1 - \Omega^2 \right) \sigma & = & g \cos{ \psi } \left\{ J_0(\sigma) - J_2(\sigma) \right\} 
\label{eq:fiftynine} \\ 
     - 2 \epsilon \, \Omega \sigma & = & g \sin{ \psi } \left\{ J_0(\sigma) + J_2(\sigma) \right\}  \: , 
\label{eq:sixty}
\end{eqnarray}
where $J_{n}(\sigma)$ is Bessel function of order $n$. Applying the identities $J_0(\sigma) - J_2(\sigma) = 2 J'_1(\sigma)$ and $J_0(\sigma) + J_2(\sigma) = 2 J_1(\sigma)/\sigma$, equations~(\ref{eq:fiftynine})-(\ref{eq:sixty}) are rewritten as
\begin{eqnarray}
\left( 1 - \Omega^2 \right) \left[ \frac{\sigma}{2 J'_1(\sigma)} \right] & = & g \cos{ \psi }  
\label{eq:sixtyonetwo} \\ 
     - 2 \epsilon \, \Omega \left[ \frac{\sigma^2}{2J_1(\sigma)} \right] & = & g \sin{ \psi } \: . 
\label{eq:sixtyonethree}
\end{eqnarray}
Squaring and summing these equations eliminates phase $\psi$:
\begin{equation}
\left( 1 - \Omega^2 \right)^2 + 
\left\{ 2 \epsilon \left[ \frac{\sigma J'_1(\sigma)}{J_1(\sigma)} \right] \Omega \right\}^2 
= \left[ \frac{2 J'_1(\sigma) g}{\sigma} \right]^2 .
\label{eq:sixtyonefour}
\end{equation}
It is convenient to introduce renormalized force and dissipation parameters:
\begin{eqnarray}
\tilde{g}_{r}        & \equiv & g        \cdot \left[ 2 J'_1(\sigma) \right] ,   
\label{eq:sixtyonefive} \\ 
\tilde{\epsilon}_{r} & \equiv & \epsilon \cdot \left[ \frac{\sigma J'_1(\sigma)}{J_1(\sigma)} \right]  . 
\label{eq:sixtyonesix}
\end{eqnarray}
The subscript $r$ here stands for ``resonance''. The most important features of $\tilde{g}_{r}$ and $\tilde{\epsilon}_{r}$ is that they tend to zero as the derivative of the $n=1$ Bessel function approaches its first zero or, equivalently, as $J_1(\sigma)$ approaches its first maximum. This occurs at $\sigma = 1.85$. At small amplitudes, $J_1(\sigma \ll 1) \approx \sigma/2$, $J'_1(\sigma \ll 1) \approx 1/2$, and $\tilde{g}_{r} \rightarrow g$, $\tilde{\epsilon}_{r} \rightarrow \epsilon$. 

\begin{figure}[t]
\begin{center}
\includegraphics[width=0.48\textwidth]{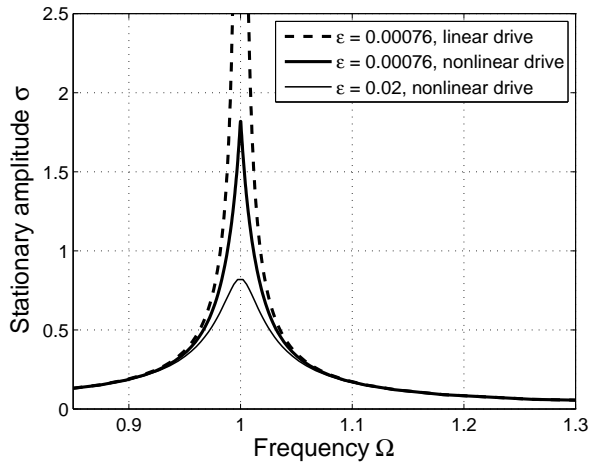}
\end{center}
\caption{The resonance curve (\ref{eq:sixtyoneseven}) corresponding to nonlinear drive and linear spring (thick and thin solid lines). The parameters are $g = 0.036$, $\delta = 0$, $\epsilon = 0.00076$ (the thick line) and $\epsilon = 0.02$ (the thin line). The dashed line is the linear drive resonance curve for $\epsilon = 0.00076$ and $g = 0.036$.}
\label{fig:fortyone}
\end{figure}

Equation (\ref{eq:sixtyonefour}) establishes a relation between the drive frequency $\Omega$ and the amplitude $\sigma$. It is easier to express $\Omega$ vs. $\sigma$ rather than vice versa:
\begin{equation}
\Omega_{\rm res. \: drive} = 
\sqrt{ \left( 1 - 2 \tilde{\epsilon}^2_{r} \right) \pm \! 
\sqrt{ \frac{\tilde{g}^2_{r}}{\sigma^2} - 4 {\tilde{\epsilon}}^2_{r} + 4 {\tilde{\epsilon}}^4_{r} }} \: .
\label{eq:sixtyoneseven}
\end{equation}
This resonance function is shown in figure~\ref{fig:fortyone} as the thick and thin solid lines. The thick line corresponds to weak damping $\epsilon = 0.00076$. At this value, the amplitude of forced oscillations increases until the renormalized force $\tilde{g}_{r}$ starts to decrease. The two processes balance each other and the stationary amplitude converges to 1.85. It must be emphasized that the amplitude would have been much larger than 1.85 in the usual case of linear driving at the same damping. Such a curve is shown in figure~\ref{fig:fortyone} by the dashed line.

One concludes that at weak damping the amplitude of forced oscillations is no longer limited by damping as in conventional linear resonance. Instead, it converges to a constant value of 1.85. In dimensional units, see equation~(\ref{eq:fortyfive}), it corresponds to  
\begin{equation}
\bar{x}_m = \frac{1.85}{2\pi} \, L \approx 0.294 \, L \: . 
\label{eq:sixtythree}
\end{equation}
Thus three-phase driving possesses an important property of self-limitation. The stationary amplitude is weakly dependent on the force magnitude $g$ and is directly proportional to the spatial period of $ABC$ array. Another property of the nonlinear force is that the stationary amplitude is weakly dependent on damping rate $\epsilon$ as long as the latter is below a critical value.  

At larger $\epsilon$ (small $Q$), damping becomes so strong that at some point it starts to limit the amplitude. It happens when the tip of the resonance curve detaches from $\sigma = 1.85$. In other words, the inner root in equation~(\ref{eq:sixtyoneseven}) becomes zero at $\sigma = 1.85$. Expanding this condition and neglecting $\tilde{\epsilon}^4_{r}$ relative to $\tilde{\epsilon}^2_{r}$, one obtains
\begin{equation}
\epsilon_{\rm cr} = g \cdot \left. \frac{J_1(\sigma)}{\sigma^2} \right\vert_{\sigma = 1.85} = 0.169 \, g \: . 
\label{eq:sixtythreeone}
\end{equation}
For $g = 0.036$, this yields $\epsilon_{\rm cr} = 0.0061$ (and the critical quality factor $Q_{\rm cr} = 82$). This is about 8 times larger than $\epsilon = 0.00076$. Thus the parameter set of table~\ref{tab:three} is well within the pitch-limiting regime. An example of damping-limiting regime, $\epsilon = 0.02$, is shown in figure~\ref{fig:fortyone} by the thin solid line. In this case, the stationary amplitude only grows to 0.82 at resonance, a value well below the pitch limit of 1.85.

\begin{figure}[t]
\begin{center}
\includegraphics[width=0.48\textwidth]{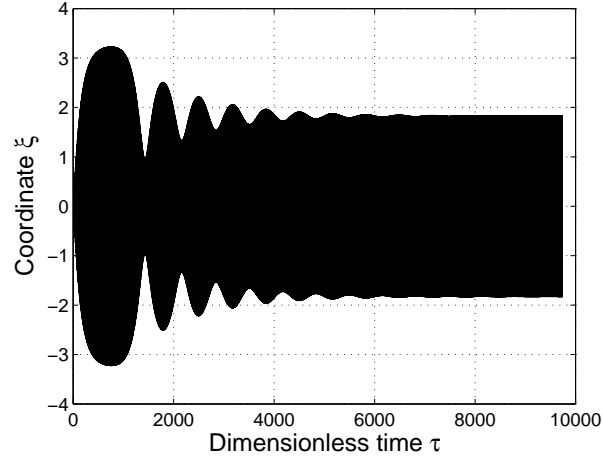}
\end{center}
\caption{Numerical solution of equation~(\ref{eq:fiftyfive}) for $\Omega = 1.0$, $\epsilon = 0.00076$ and $g = 0.036$. After swings oscillations, the amplitude settles at 1.85. The initial conditions are $\xi(0) = \xi'(0) = 0$. }
\label{fig:four}
\end{figure}

\begin{figure}[t]
\begin{center}
\includegraphics[width=0.48\textwidth]{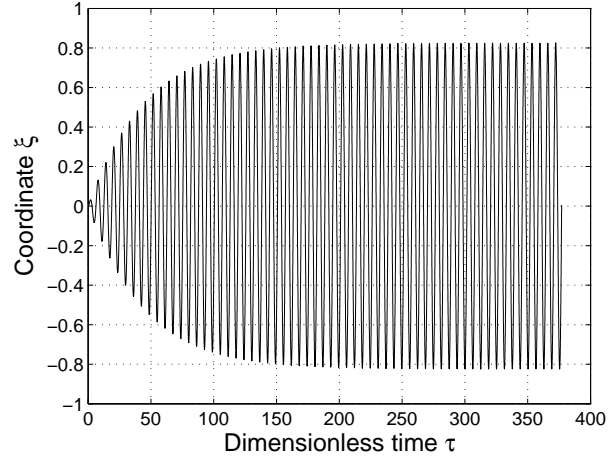}
\end{center}
\caption{Numerical solution of equation~(\ref{eq:fiftyfive}) for $\Omega = 1.0$, $\epsilon = 0.02$ and $g = 0.036$. The amplitude gradually approaches the stationary value of 0.82 with no swings. The initial conditions are $\xi(0) = \xi'(0) = 0$.}
\label{fig:eight}
\end{figure}

The above analysis is supported by direct solutions of the equation of motion (\ref{eq:fiftyfive}). Figures~\ref{fig:four} and \ref{fig:eight} compare time evolution of the translator position in the pitch-limited and damping-limited regimes. In figure~\ref{fig:four}, damping is weak, $\epsilon = 0.00076$, and motion is underdamped. Initially, the amplitude grows fast and quickly outgrows the stationary value of 1.85. Then the driving force effectively changes sign and becomes a stopping force. As a result, the amplitude drops below 1.85. Such swing cycles continue for some time until their intensity subsides and the amplitude settles at the stationary value. In contrast, in figure~\ref{fig:eight}, $\epsilon = 0.02$, and the motion is overdamped. No amplitude swings are observed. Instead, the amplitude rises monotonically and smoothly approaches a stationary value of 0.82.

\section{\label{sec:seven}
Parametric driving
}

In resonant driving considered in the preceding section, the external force was set in phase with translator velocity $\dot{\bar{x}}$ which led to the phase condition (\ref{eq:forty}) and the equation of motion (\ref{eq:fiftythree}). This condition maximizes the power transferred to the oscillator. In parametric driving, the external force is set in phase with translator displacement ${\bar x}$. That is, the main driving force is zero when ${\bar x} = 0$. From this and equation~(\ref{eq:thirtynine}) follows a different, parametric three-phase condition 
\begin{equation}
k_{1} x_0 + \theta = 0 \, , \; \pi   \:\: .
\label{eq:sixtyfour}
\end{equation}
The main driving force is [the last term in equation~(\ref{eq:thirtythree})]: 
\begin{equation}
\pm \, 3 k_1 V U G^{(7)}_{1} \sin{ ( k_{1} \bar{x} ) } \cos{ \omega t } \: . 
\label{eq:sixtyfive}
\end{equation}
The overall sign is plus or minus depending on whether zero or $\pi$ is chosen. As a matter of fact, the choice is irrelevant since the sign can be flipped by shifting the time origin by $\pi/\omega$. It is convenient to choose the negative sign. The $V^2$ term in the truncated force (\ref{eq:thirtythree}) is defined by the factor
\begin{equation}
\sin{ ( k_{2} x + 2 \theta ) }  = \sin{ [ k_{2} \bar{x} + 2 ( k_1 x_0 + \theta ) ] } = 
\sin{ ( k_{2} \bar{x} + \{ 0, (2\pi) \} ) } = \sin{ ( k_{2} \bar{x} ) } \: .  
\label{eq:sixtysix}
\end{equation}
Substituting the truncated force into the full equation of motion and switching to dimensionless units as in section~\ref{sec:six} one obtains   
\begin{equation}
\xi'' + 2 \epsilon \xi' + \xi  
- \delta \sin{ ( 2\xi ) } + g \sin{ ( \xi ) } \cos{ \Omega \tau } = 0 . 
\label{eq:sixtyseven}
\end{equation}
Comparing to resonant driving, this equation of motion has a different sign of the $\delta$ term and a sine instead of cosine nonlinearity in the driving force. The $\sin{\xi} \cos{\Omega t}$ nonlinearity has been extensively studied over decades in relation to parametrically excited pendulum~\cite{Kapitsa1951a,Landau1976,McLaughlin1981,Leven1981,Croquette1981,Koch1983,Koch1985,Leven1985,Meissner1986,Briggs1987,Kim1996,Kwek1996,Kim1997,Blackburn2006}. Again, the system under study is more general. Its elastic force is decoupled from the driving force and as such can be of arbitrary form allowed by the physics of elastic springs.  

The method of slowly changing amplitudes can also be applied to parametric driving. Limiting again consideration to $\delta = 0$, the equation of motion (\ref{eq:sixtyseven}) reduces to  
\begin{equation}
\xi'' + 2 \epsilon \xi' + \xi + g \sin{ ( \xi ) } \cos{ \Omega \tau } = 0 \: . 
\label{eq:seventy}
\end{equation}
The solution is sought in a single-harmonic form
\begin{equation}
\xi(\tau) = \sigma(\tau) \cos{ \left[ \omega \tau + \psi(\tau) \right] } \: , 
\label{eq:seventyone}
\end{equation}
where $\omega \approx 1$ is close to the natural frequency of the proof mass. The external frequency is set to $\Omega = 2 \omega$. In the limit of stationary oscillations, $\sigma$ are $\psi$ are treated as constants. Substituting in equation~(\ref{eq:seventy}) and shifting the time origin by $\psi/\omega$ yields 
\begin{eqnarray}
\left( 1 - \omega^2 \right) \sigma \cos{\omega \tau} 
- 2 \epsilon \omega \sigma \sin{\omega \tau} = 
\nonumber \\ 
\makebox[1.0cm]{}
g \cos{(2\psi)} \sin{ \left\{ \sigma \cos{\omega \tau} \right\} } \cos{( 2 \omega \tau )}  + 
g \sin{(2\psi)} \sin{ \left\{ \sigma \cos{\omega \tau} \right\} } \sin{( 2 \omega \tau )} \: .   
\label{eq:seventytwo}
\end{eqnarray}
Multiplying by $\cos{(\omega \tau)}$ and averaging over time, then multiplying by $\sin{(\omega \tau)}$ and averaging, and finally applying the identities $J_1(\sigma) - J_3(\sigma) = 2J'_2(\sigma)$ and $J_1(\sigma) + J_3(\sigma) = 4J_2(\sigma)/\sigma$, one obtains
\begin{eqnarray}
\sigma \left( 1 - \omega^2 \right) & = & \left[ 2 J'_2(\sigma) \right] g \cos{(2\psi)} ,
\label{eq:seventyseven} \\
- 2 \epsilon \omega \sigma                            & = & 
\!\! \left[ \frac{4 J_2(\sigma)}{\sigma} \right] g \sin{(2\psi)} . 
\label{eq:seventythree}
\end{eqnarray}
By introducing new renormalized force and dissipation parameters
\begin{eqnarray}
\tilde{g}_{p}          & \equiv & g \cdot \left[ 2 J'_2(\sigma) \right] 
= g \cdot \left[ J_1(\sigma) - J_3(\sigma) \right] \: ,
\label{eq:seventyfive} \\
\tilde{\epsilon}_{p}   & \equiv & \epsilon \cdot \left[ \frac{\sigma J'_2(\sigma)}{2 J_2(\sigma)} \right] 
= \epsilon \cdot \frac{J_1(\sigma)-J_3(\sigma)}{J_1(\sigma)+J_3(\sigma)} \: , 
\label{eq:seventysix}
\end{eqnarray}
where subscript $p$ stands for ``parametric'', the equations are cast in the form
\begin{eqnarray}
\sigma \left( 1 - \omega^2 \right)         & = & \tilde{g}_{p} \cos{(2\psi)} \: ,
\label{eq:seventyeight} \\
- 2 \tilde{\epsilon}_{p} \: \omega \sigma  & = & \tilde{g}_{p} \sin{(2\psi)} \: . 
\label{eq:seventynine}
\end{eqnarray}
This is the same functional form as equations~(\ref{eq:sixtyonetwo}) and (\ref{eq:sixtyonethree}) but with $g$ and $\epsilon$ replaced with $\tilde{g}_{p}$ and $\tilde{\epsilon}_{p}$. Accordingly, the resonance curve is given by the function similar to equation~(\ref{eq:sixtyoneseven}):
\begin{equation}
\omega_{\rm param. \: drive} = \sqrt{ \left( 1 - 2 \tilde{\epsilon}^2_{p} \right) \pm 
\sqrt{ \frac{\tilde{g}^2_{p}}{\sigma^2} - 4 \tilde{\epsilon}^2_{p} + 4 \tilde{\epsilon}^4_{p} }} .
\label{eq:seventyten}
\end{equation}
Both $\tilde{g}_{p}$ and $\tilde{\epsilon}_{p}$ tend to zero at the first maximum of the $n = 2$ Bessel function, which occurs at $\sigma = 3.05$. By analogy with the resonant case, one expects that 3.05 will be the maximal possible stationary amplitude of parametric drive. In real units 
\begin{equation}
\bar{x}_m = \frac{3.05}{2\pi} \, L \approx 0.485 \, L \: . 
\label{eq:seventyfour}
\end{equation}
One concludes that parametric driving results in about 65\% larger amplitude than resonant driving (0.485 $L$ vs. 0.294 $L$).

\begin{figure}[t]
\begin{center}
\includegraphics[width=0.48\textwidth]{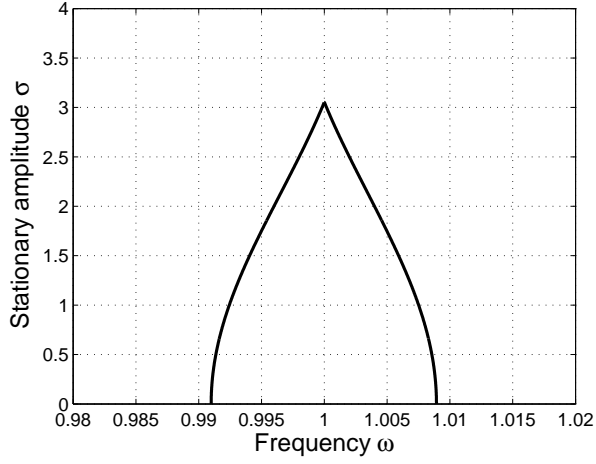}
\end{center}
\caption{Parametric resonance curve (\ref{eq:seventyten}) for $\epsilon = 0.00076$ and $g = 0.036$. External driving frequency is $\Omega = 2 \omega$.}
\label{fig:five}
\end{figure}
\begin{figure}[t]
\begin{center}
\includegraphics[width=0.48\textwidth]{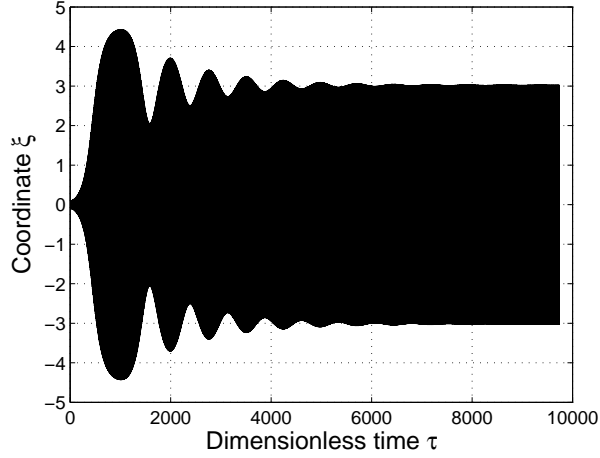}
\end{center}
\caption{Numerical solution of equation~(\ref{eq:sixtyseven}) for $\epsilon = 0.00076$, $\delta = 0$, $g = 0.036$, and $\Omega = 2.0$. Initial conditions are $\xi(0) = 0$, $\xi'(0) = 0.1$. The stationary amplitude is 3.05.}
\label{fig:nine}
\end{figure}

Another important feature of $\tilde{g}_{p}$ is that it tends to zero at small $\sigma \ll 1$ as $\propto (\sigma/2)$. Accordingly, the ratio $\tilde{g}_{p}/\sigma$ tends to a constant at small amplitudes. As a result, stationary oscillations exist only within a small frequency interval around the natural frequency $\omega = 1$. This is in sharp contrast with resonant driving. The resonance curve (\ref{eq:seventyten}) is plotted in figure~\ref{fig:five}. Its shape is distinctly different from that of resonant driving. Parametric oscillations exist within a small frequency window. The self-limitation property of parametric drive is confirmed by direct solution of the equation of motion (\ref{eq:sixtyseven}). Figure~\ref{fig:nine} shows time evolution of the translator amplitude in the underdamped regime, $\epsilon = 0.00076$. The amplitude exhibits slowly varying swings, similar to figure~\ref{fig:four}, before settling at a stationary value of 3.05, in accordance with equation~(\ref{eq:seventyfour}).

\section{ \label{sec:ten}
Summary
}

In this paper, an excitation method for MEMS devices with planar electrodes has been presented. The key ingredient is the periodic electrostatic profile generated by electrode array $ABCABC$. By independently adjusting voltages at $A$, $B$, and $C$, one can arbitrarily move the potential profile along the $x$-axis thereby compensating any fabrication misalignment between the stationary and moving parts of the device. A time dependent voltage applied to the translator array $ababab$ interacts with the stationary profile and either drives the translator into resonance or induces parametric excitation. 

Using planar electrodes to induce sliding motion necessarily involves fringe fields. The underlying electrostatic problem is complex and requires careful analysis. Various symmetries of the capacitance matrix have been analyzed and summarized in table~\ref{tab:one}. The capacitance coefficients have been computed by a finite-element numerical method as functions of the translator position for a realistic electrode geometry. The results are shown in figure~\ref{fig:onetwo}. Fourier coefficients are given in table~\ref{tab:two}. General expressions for quasi-electrostatic energy and force have been derived in sections~\ref{sec:three} and \ref{sec:four}, respectively. Based on the analysis of Fourier coefficients, a truncated electrostatic force (\ref{eq:thirtythree}) has been derived. In addition to the expected main driving term, the force contains a stator-induced self-force that affects the translator dynamics. The basic equation of translator motion has been derived for both resonant driving, equation~(\ref{eq:fiftythree}), and parametric driving, equation~(\ref{eq:sixtyseven}). 

Spatial periodicity of the electrostatic potential results in a periodic dependence of the driving force on the translator displacement. It leads to self-stabilization of forced oscillations. For weak dissipation, the amplitude of forced oscillation stabilizes at 0.294 of the electrode array pitch in the case of resonant driving, and at 0.485 of the array pitch in the case of parametric driving. Self-stabilization has been confirmed by direct solution of the equations of motion in time domain. 

Self-stabilization can be useful in applications where amplitude stability is required, for instance, in high-performance vibratory MEMS gyroscopes. Temperature, pressure and other environmental variations do not affect the array pitch to the same degree as they affect quality factor $Q$. Pitch limited nonlinear excitation methods may provide a higher degree of stability than dissipation limited ones.

\begin{acknowledgments}

The authors wish to thank Rod Alley, Vasiliy Baydulov, Robert Bicknell, Sergej Flach, Peter Hartwell, Brian Homeijer, Matt Hopcroft, Bernardo Huberman, Richard Martin, Don Milligan, Peter Nyholm, Jeremy Sells, Wesley Smith and Oleg Yevtushenko for numerous discussions on the subject of this paper, Tatiana Kornilovich for help with references, and Kenneth Abbott, Chris Davis, Kenneth Vandehey, and Timothy Weber for supporting this work.      

\end{acknowledgments}

\section*{References}

\providecommand{\noopsort}[1]{}\providecommand{\singleletter}[1]{#1}%

\end{document}